\documentstyle[prd,aps,epsfig]{revtex}

\newcommand{\be}{\begin{equation}}
\newcommand{\ee}{\end{equation}}

\begin{document}
\title{Systematics of $q\bar q$-states in the
$(n,M^2)$ and $(J,M^2)$ planes}
\author{A.V. Anisovich, V.V. Anisovich, and A.V. Sarantsev}
\address{St.Petersburg Nuclear Physics Institute, Gatchina, 188350,
Russia}
\date{March 13, 2000}
\maketitle

\begin{abstract}
In the mass region up to $M<2400$ MeV
we systematise mesons on the plots $(n,M^2)$ and $(J,M^2)$,
thus setting their classification
in terms of $n\; ^{2S+1}L_J \; q\bar q$ states.
The trajectories on the $(n,M^2)$--plots are
drawn for the following $(IJ^{PC})$-states:
$a_0(10^{++})$, $a_1(11^{++})$,
$a_2(12^{++})$, $a_3(13^{++})$, $a_4(14^{++})$,
$\pi(10^{-+})$, $\pi_2(12^{-+})$,
$\eta(00^{-+})$, $\eta_2(02^{-+})$, $\rho(11^{--})$, $f_0(00^{++})$,
$f_2(02^{++})$.
All trajectories are linear, with nearly the same slopes. At
the  $(J,M^2)$--plot we set out meson states
for leading and daughter
trajectories: for
$\pi$, $\rho$, $a_1$, $a_2$ and $P'$.

\end{abstract}

\bigskip

In the last decade tremendous efforts have been paid to study
meson spectra over the mass region 1000--2400 MeV. The collected
rich information that includes the discovery of new resonances and
confirmation of those discovered before needs to be systematized.

We present a scheme for  $q\bar q$ trajectories on the
$(n,M^2)$ and $(J,M^2)$ plots ($n$ is the radial quantum number and
$J$ is the meson spin)
using the latest results \cite{pif2,a1(1640),pieta,eta,WA,NuclPhys}
together with previously accumulated data  \cite{PDG}.

The trajectories on the  $(n,M^2)$--plots are presented in Figs. 1
and 2:
they are linear and  with a good accuracy can be represented as
\be
M^2=M_0^2+(n-1)\mu^2.
\ee
$M_0$ is the mass of basic meson and $\mu^2$ is the
trajectory slope parameter: $\mu^2$ is approximately the same for all
trajectories: $\mu^2 = 1.25\pm 0.15$ GeV$^2$.

At $M\le 2400$ MeV the
mesons of  $q\bar q$ nonets $n\; ^{2S+1}L_J$
fill in the $(n, M^2)$-trajectories as follows:
\begin{equation} \begin{array}{lll}
^1S_0 \to \pi(10^{-+}),\; \eta(00^{-+})\ , &
^3S_1 \to \rho(11^{--}),\; \omega(01^{--})\ ; &\\
^1P_1 \to b_1(11^{+-}),\; h_1(01^{+-}), &
^3P_J \to a_J(1J^{++}),\; f_J(0J^{++}) &  J=0,1,2\ ;\\
^1D_2 \to \pi_2(12^{-+}),\; \eta_2(02^{-+}), &
^3D_J \to \rho_J(1J^{--}),\; \omega_J(0J^{--}) &
J=1,2,3\  ;\\
^1F_3 \to b_3(13^{+-}),\; h_3(03^{+-}), &  ^3F_J \to
a_J(1J^{++}),\; f_J(0J^{++}) &  J=2,3,4\ .
\end{array}
\end{equation}
Trajectories with the
same $IJ^{PC}$ can be created by different orbital momenta with
$J=L\pm 1$, in this way they are doubled: these are trajectories
$(I1^{--})$, $(I2^{++})$, and so on.

Isoscalar states are formed by two light flavor components,
$n\bar n=(u\bar u+d\bar d)/\sqrt 2$ and $s\bar s$. Likewise, this
also results in doubling  isoscalar trajectories.

The trajectories $a_1(11^{++})$ and $a_3(13^{++})$
are shown in Fig. 1a:
the states $a_1(2100)$, $a_1(2340)$, $a_3(2070)$, $a_3(2310)$ have been
seen in \cite{pif2}, an evidence for $a_1(1640)$ was found in
\cite{a1(1640)}.

Figure 1b displays the trajectories $\eta(00^{-+})$ and
$\eta_2(02^{-+})$: the doubling is due to
two independent flavor components $n\bar n$ and $s\bar s$. The
states  $\eta_2(2030)$ and $\eta_2(2300)$
have been seen in the analysis \cite{eta}. Linear extrapolation of
trajectories predicts the mesons
$\eta(1900)$, $\eta(2100)$ and $\eta_2(2200)$.

Two trajectories $\rho(11^{--})$ related to $S$
and $P$ $q\bar q$-waves are
demonstrated at Fig. 1c. The states $\rho(1700)$ and
$\rho(2150)$ are cited in \cite{PDG}, the trajectories
predict  $\rho(1830)$, $\rho(2060)$ and $\rho(2380)$.

In Fig. 1d one can see the trajectories
for  $10^{-+}$ and  $12^{-+}$. The states
$\pi(1300)$, $\pi(1800)$
$\pi_2(1670)$ are from \cite{PDG}; all other states are predicted by
the linearity of trajectories.

Experimental data in the sector $a_0(10^{++})$, $a_2(12^{++})$,
$a_4(14^{++})$ do not fix the slope $\mu^2$ uniquely.
In Fig. 2a the trajectories $a_0(10^{++})$, $a_2(12^{++})$
and $a_4(14^{++})$ are shown
for $\mu^2 =1.38$ GeV$^2$. The states
$a_0(2000^{+50}_{-100})$, $a_2(1980)$, $a_2(2100)$, $a_2(2280)$,
$a_4(2005)$,  $a_4(2260)$ have
been observed in  \cite{pif2,pieta}
and $a_2(1640\pm 60)$ in \cite{AKP}. Two trajectories
$a_2(12^{++})$  owe their existence to two states,
$^3P_2 q\bar q$ and $^3F_2 q\bar q$. Obviously the upper one  refers
to states with dominant $^3F_2 q\bar q$ component.
In Fig. 2b one can see the
trajectories $a_0(10^{++})$, $a_2(12^{++})$,
$a_4(14^{++})$ for $\mu^2=1.1$ GeV$^2$.
Comparison of Figs. 2a and 2b demonstrates the
uncertainty in fixing  $a_J$-resonances.
A noticeable difference
between Figs. 2a and 2b consists in the prediction of $a_0(1800)$ for
$\mu^2=1.1$ GeV$^2$.

For the $02^{++}$ states a quadruplet of trajectories arises due to
the presence of both two waves, $^3P_2$ and $^3F_2$, and two
flavor components, $n\bar n$ and $s\bar s$.
To set $f_2(02^{++})$ mesons on  $(n, M^2)$ trajectories
with $\mu^2 =1.38$ GeV$^2$
faces a problem, that is seen in
Fig. 2b:  we cannot fit the experimental data
unambiguously.
We know definitely two $1^3P_2 q\bar q$ states
which are $f_2(1285)$ and
$f_2(1525)$. This establishes
masses of other $f_2$ mesons on these trajectories: they
are to be near $ 1700$ MeV, $ 1940$ MeV, $ 2060 $ MeV,
$ 2260 $ MeV, $ 2390 $ MeV. The trajectories of $1^3F_2 q\bar q$ states
are located higher, with a gap of the order of $\Delta M^2 \simeq 2.5 $
GeV$^2$, that gives mesons with masses around $2050$ MeV,
$2200$ MeV,  $2350$ MeV. The compilation \cite{PDG} presents three
candidates for $f_2$ with the mass near $1700$ MeV:
$f_2(1640)$, $f_J(1710)$, and $f_2(1800)$; the resonance $f_2(1950)$
has been seen in $\pi\pi\pi\pi$ \cite{WA}. Simultaneous analysis of the
$\pi\pi$, $\eta\eta$  and $\eta\eta'$ spectra \cite{NuclPhys} gives
evidence for $f_2(1910)$, $f_2(2020)$, $f_2(2230)$ and $f_2(2300)$;
in the compilation \cite{PDG},
the state $f_2(2150)$ is also under discussion. In
the $\phi\phi$ spectra three tensor resonances were seen:
$f_2(2010)$, $f_2(2300)$, $f_2(2340)$ \cite{BNL}.
Comparison of the predictions given by Fig. 2c with the observed
mesons does not present strong arguments in favor of $\mu^2 \simeq
1.38$ GeV$^2$. The description of data with $\mu^2 \simeq 1.10$ GeV$^2$
looks much more reasonable, see Fig. 2d.  Nevertheless, we must admit
that location of  all cited $f_2$-mesons on the trajectories is
questionable at present level of knowledge: the main problem is to
distinguish between different states with close masses. To this aim a
more sophisticated technique is needed, that is, a simultaneous analysis
of available data in the framework of the $K$-matrix approach or $N/D$
method.

In Fig. 2e the trajectories $f_0(00^{++})$ are displayed:
they are doubled due to two flavor components,
$n\bar n$ and $s\bar s$. We do not put the
enigmatic $\sigma$-meson \cite{Narison,MP,LM,AN} on the $q\bar q$
trajectory
supposing $\sigma$ is alien to this classification.  The broad
state $f_0(1530^{+90}_{-250})$ (or $\epsilon (1400)$ in old notation),
which is the descendant of  the scalar glueball after mixing with
the neighboring $q\bar q$ states \cite{K,lock}, is superfluous for the
$q\bar q$ trajectories and is also not put on the trajectory.

Trajectories $f_0(00^{++})$ in Fig. 2e are
drawn for the masses of
real resonances. However, in case of scalar/isoscalar $q\bar q$-states
there is an effect which is specific for  $00^{++}$ wave,
that is, a strong mixing of $q\bar q$ states due to their overlapping:
the transitions $resonance \;1 \to real\;mesons \to resonance \;2$
result in a considerable mass shift (detailed discussion may be found
in \cite{K,lock}). The states "before mixing" which respond to
$K$-matrix poles  were found
in \cite{K,lock} for two multiplets, $1^3P_0$ and $2^3P_0$
(these states were denoted as $f^{bare}_0$). It is
rather instructive to see the location of $f^{bare}_0$ at the
$(n,M^2)$ plane: corresponding trajectories are shown in Fig.
2f. One can see a degeneration of states belonging to
two different trajectories, just this degeneration has enlarged the
mixing of $f_0$ states in the region 1000-1800 MeV. The linearity
of trajectories tells us that a strong mixing of
scalar/isoscalar states is very possible at higher masses  as
well.

The trajectories of the $(n,M^2)$ plots should be
complemented by those of
$(J,M^2)$ plots: they are shown in Fig. 3.
The important point is the leading meson trajectories
($\pi$, $\rho$, $a_1$, $a_2$ and $P'$) are known from the study
of hadron diffractive processes at $p_{lab} \sim 5-50$ GeV/c (for example,
see \cite{regge}).

The  $\pi$-meson
trajectories, leading and daughter ones (see Fig. 3a), are linear.
The leading trajectory includes $\pi(140)$, $\pi_2(1670)$ and
$\pi_4(2350)$, while the daughter one contains
$\pi(1300)$ and $\pi_2(2100)$. The other leading trajectories
($\rho$, $\eta$, $a_1$, $a_2$, $f_2$ or $P'$)  are also compatible
with linear-type behavior:
\be
\alpha_X(M^2)\simeq\alpha_X(0)+\alpha\;'_X(0)M^2.
\ee
Parameters for leading trajectories, which are defined
by the positions of $q\bar q$ states, are as follows:
\begin{equation} \begin{array}{lll}
\alpha_\pi(0)\simeq -0.015\ , & \alpha'_\pi(0)\simeq 0.72
&\mbox {GeV}^{-2}\ ;\\
\alpha_\rho(0)\simeq 0.50\ , &\alpha'_\rho(0)\simeq 0.83
&\mbox {GeV}^{-2}\ ;\\
\alpha_\eta(0)\simeq -0.24\ , &\alpha'_\eta(0)\simeq 0.80
&\mbox {GeV}^{-2}\ ;\\
\alpha_{a_1}(0)\simeq  0\ ,&\alpha'_{a_1}(0)\simeq 0.72
&\mbox  {GeV}^{-2}\ ;\\
\alpha_{a_2}(0)\simeq 0.45\ , &\alpha'_{a_2}(0)\simeq 0.91
&\mbox {GeV}^{-2}\ ;\\
\alpha_{P'}(0)\simeq 0.71\ , &\alpha'_{P'}(0)\simeq 0.83
&\mbox {GeV}^{-2}.
\end{array}
\end{equation}

The slopes are nearly the same for all trajectories, and the
inverse value of a universal slope, $1/\alpha'_X\simeq 1.25\pm 0.15$
GeV$^{2}$, is approximately equal to the slope $\mu^2$ for trajectories
on the $(n,M^2)$--plot: $\mu^2 \simeq 1/\alpha'_X$.

All daughter trajectories for $\pi$ (Fig. 3a), $a_1$ (Fig. 3b), $\rho$
(Fig. 3c), $a_2$ (Fig. 3d) and $\eta$ (Fig. 3e) are uniquely
determined.
In the classification of daughter
$P'$ trajectories we have used the variant which corresponds to
Fig. 2d.

Concerning  $P'$-trajectories (Fig. 3f), one should notice
that along the line $0^{++}$ in the region $M\sim
1500-2000$ the density of $q\bar q$-states is lower. This is
affected by the presence of the light scalar glueball.
In this region the state $f_0(1530^{+90}_{-250})$  exists which is
the glueball descendant created after mixing with the nearest
$q\bar q$ neighbors \cite{K,lock}. Because of that, the lower density of
$00^{++}q\bar q$-levels near 1500--1800 MeV does not look eventual: the
extra state (gluonium) being mixed with $q\bar q$-state repulses
the $q\bar q$ levels.

We conclude: meson states fit to linear trajectories at the
$(n,M^2)$ and $(J,M^2)$ plots with sufficiently
good accuracy.
The linear behavior of $q\bar q$ trajectories at large $M^2$
should facilitate the $q\bar q$ systematizing which is a necessary step
in identification of extra (exotic) states.

We thank Ya.I. Azimov, D.V. Bugg, L.G. Dakhno, G.S. Danilov, S.A.
Kudryavtsev, V.A. Nikonov and A.A. Yung for useful discussions. This
work was supported by the RFFI grant 98-02-17236.


\newpage
\begin{figure}
\centerline{\epsfig{file=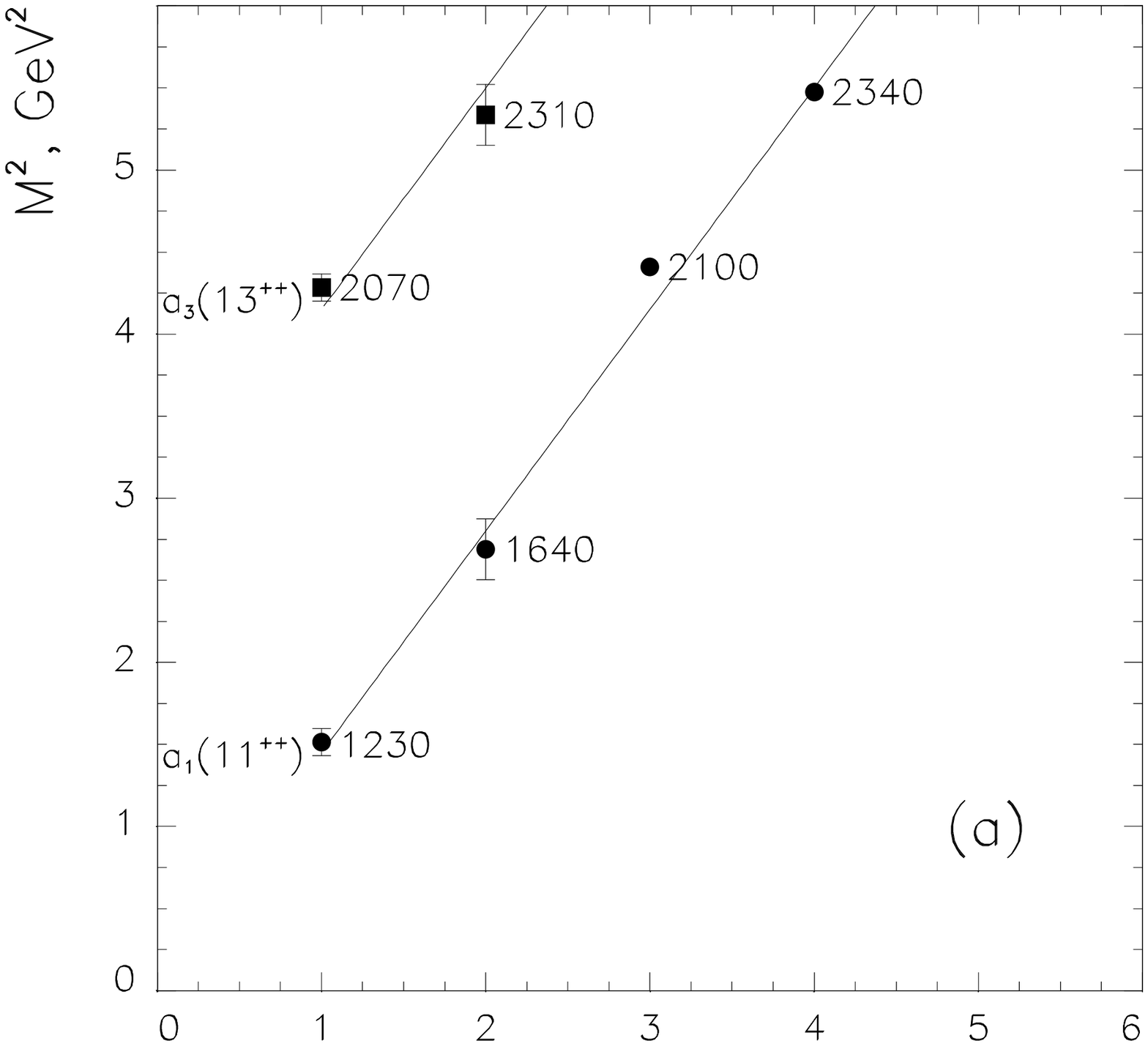,width=10cm}\hspace{-1.8cm}
            \epsfig{file=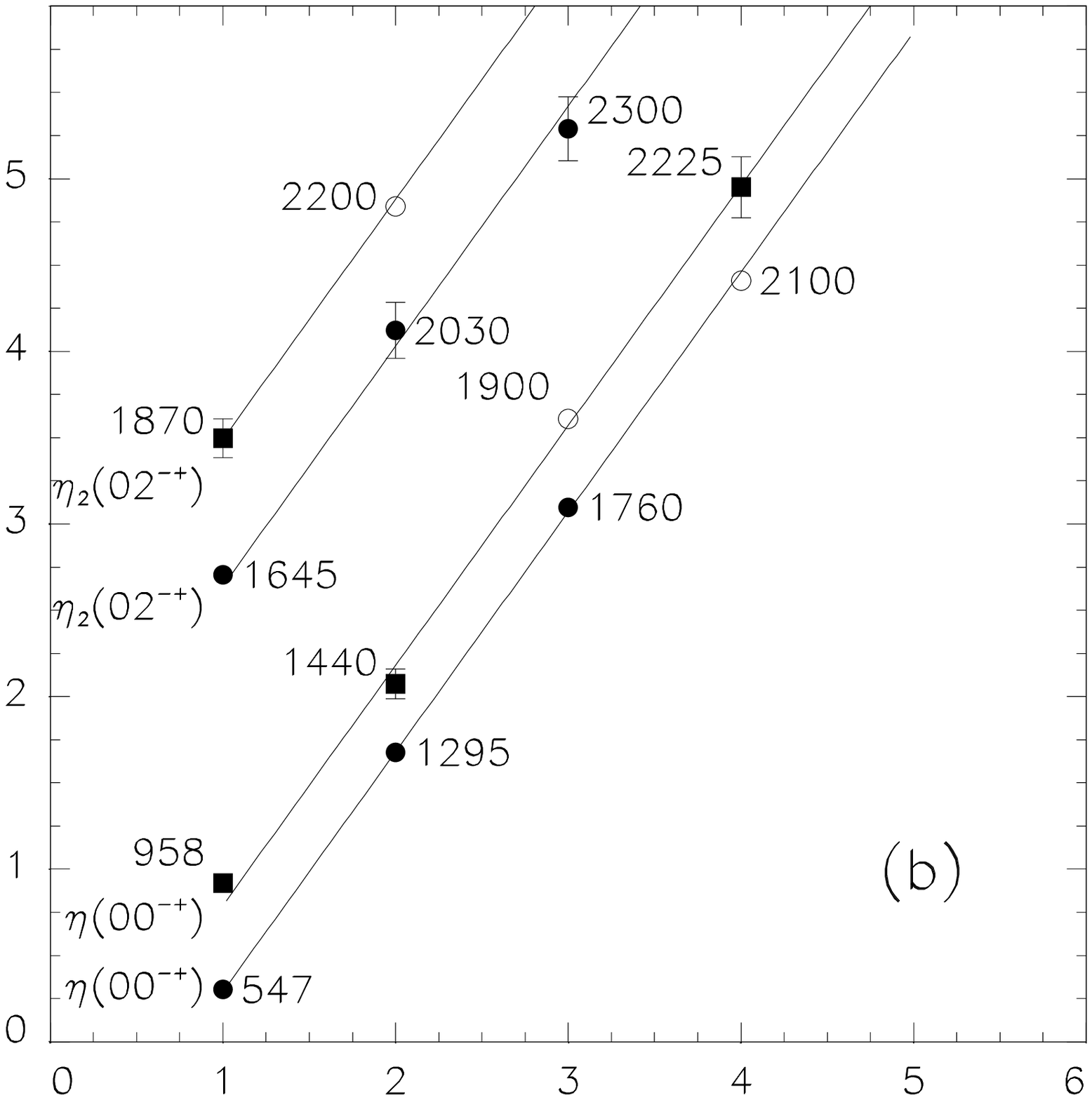,width=10cm}}
\vspace{-1.8cm}
\centerline{\epsfig{file=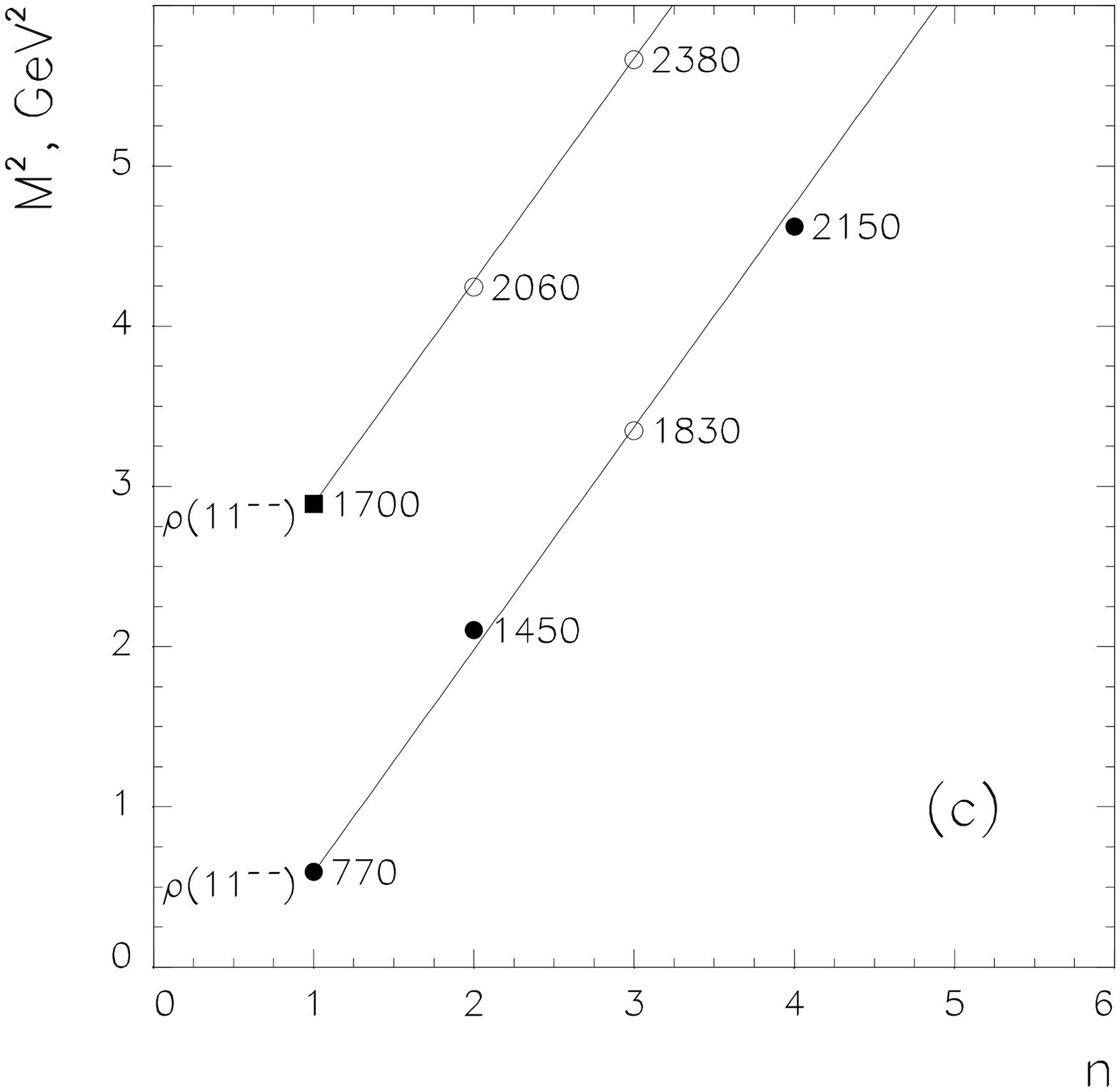,width=10cm}\hspace{-1.8cm}
            \epsfig{file=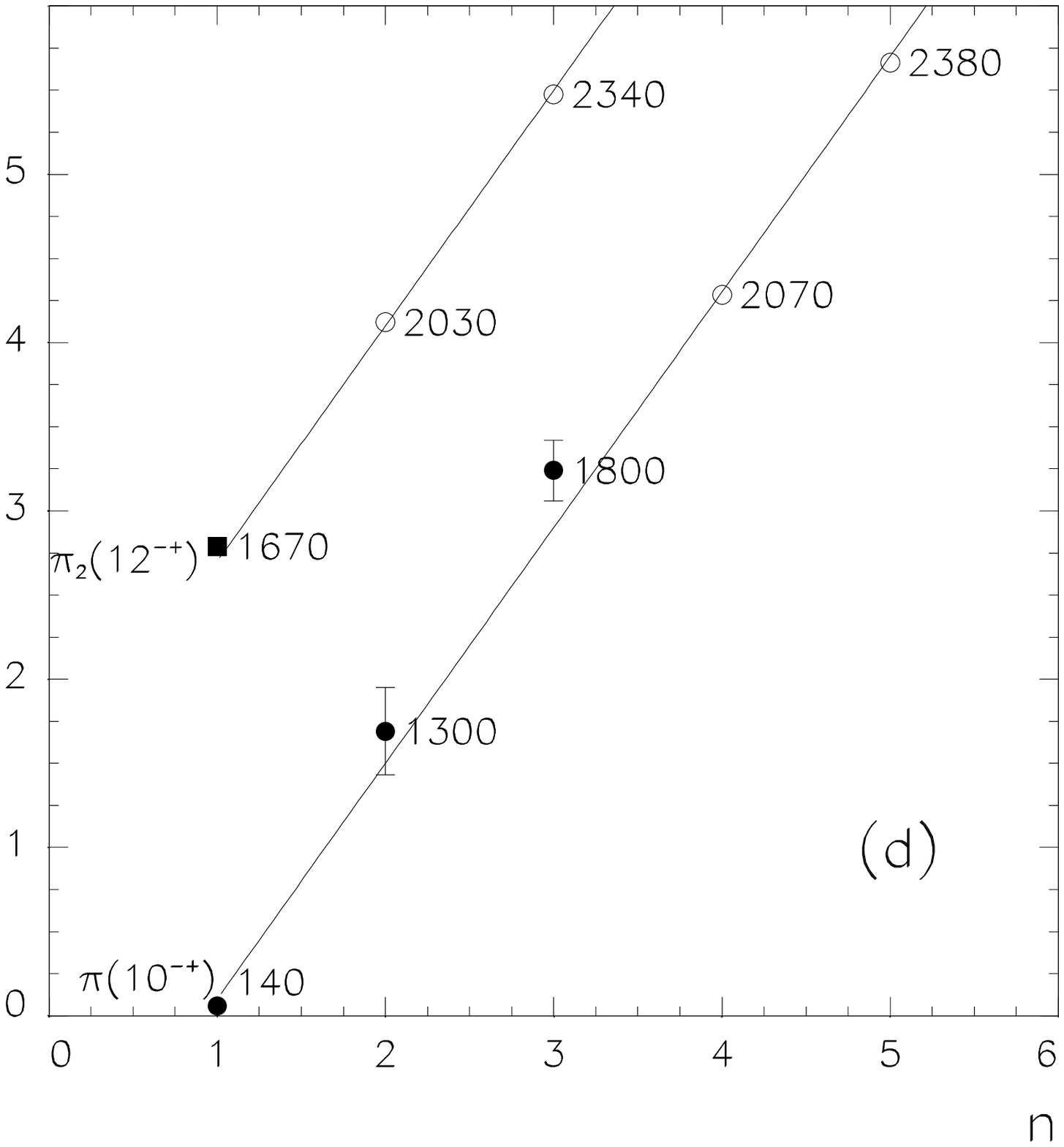,width=10cm}}
\caption{
The $(n, M^2)$-plots for the states:
a) $a_1(11^{++})$ and $a_3(13^{++})$;
b) $\eta(00^{-+})$ and $\eta_2(02^{-+})$;
c) $\rho(11^{--})$;
d) $\pi(10^{-+})$ and $\pi_2(12^{-+})$. The trajectory
slope is equal to $\mu^2=1.39$ GeV$^2$.
Open circles stand for states predicted by the present
classification.}
\end{figure}

\begin{figure}
\centerline{\epsfig{file=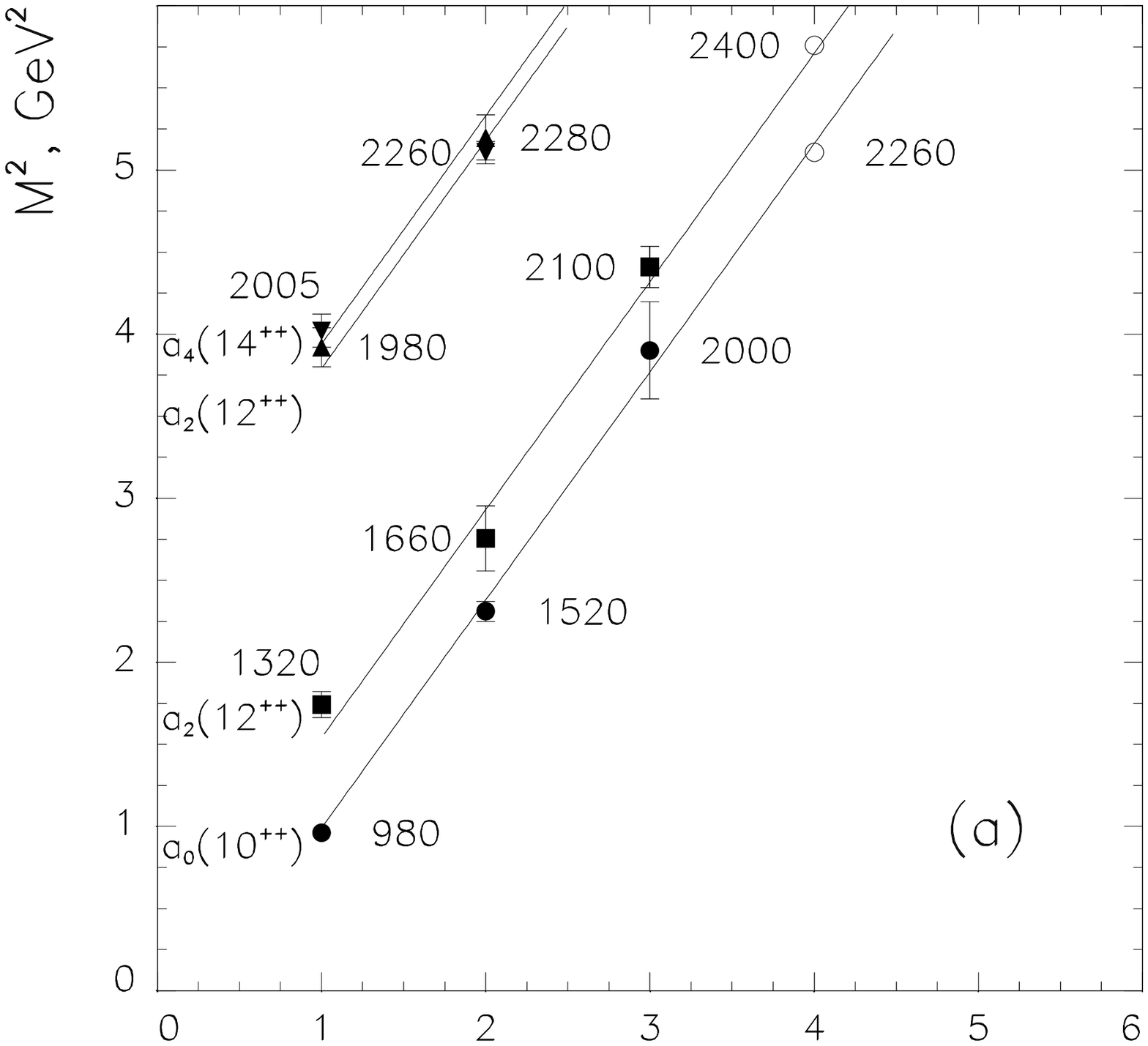,width=8cm}\hspace{-1.5cm}
            \epsfig{file=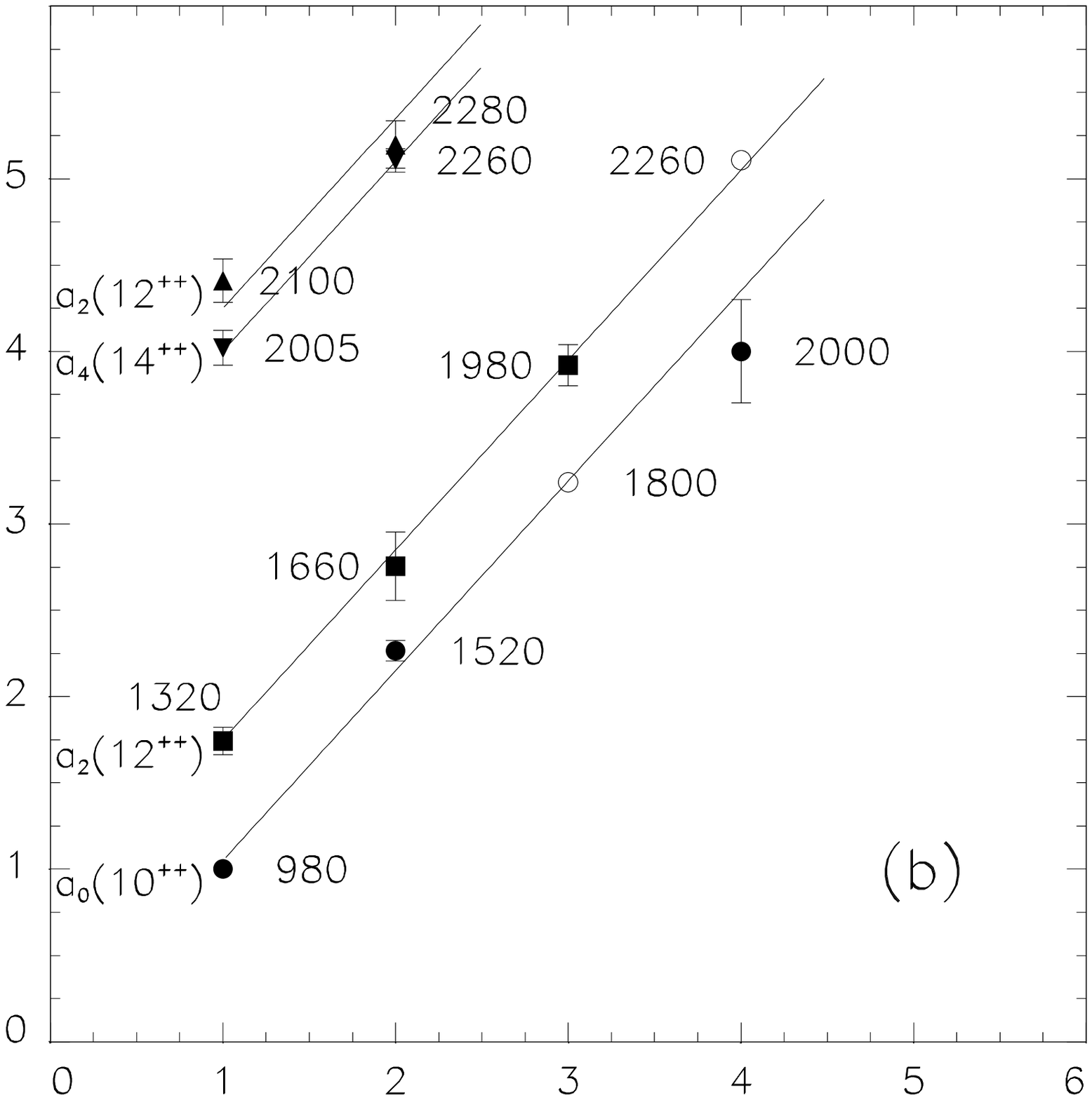,width=8cm}}
\vspace{-1.5cm}
\centerline{\epsfig{file=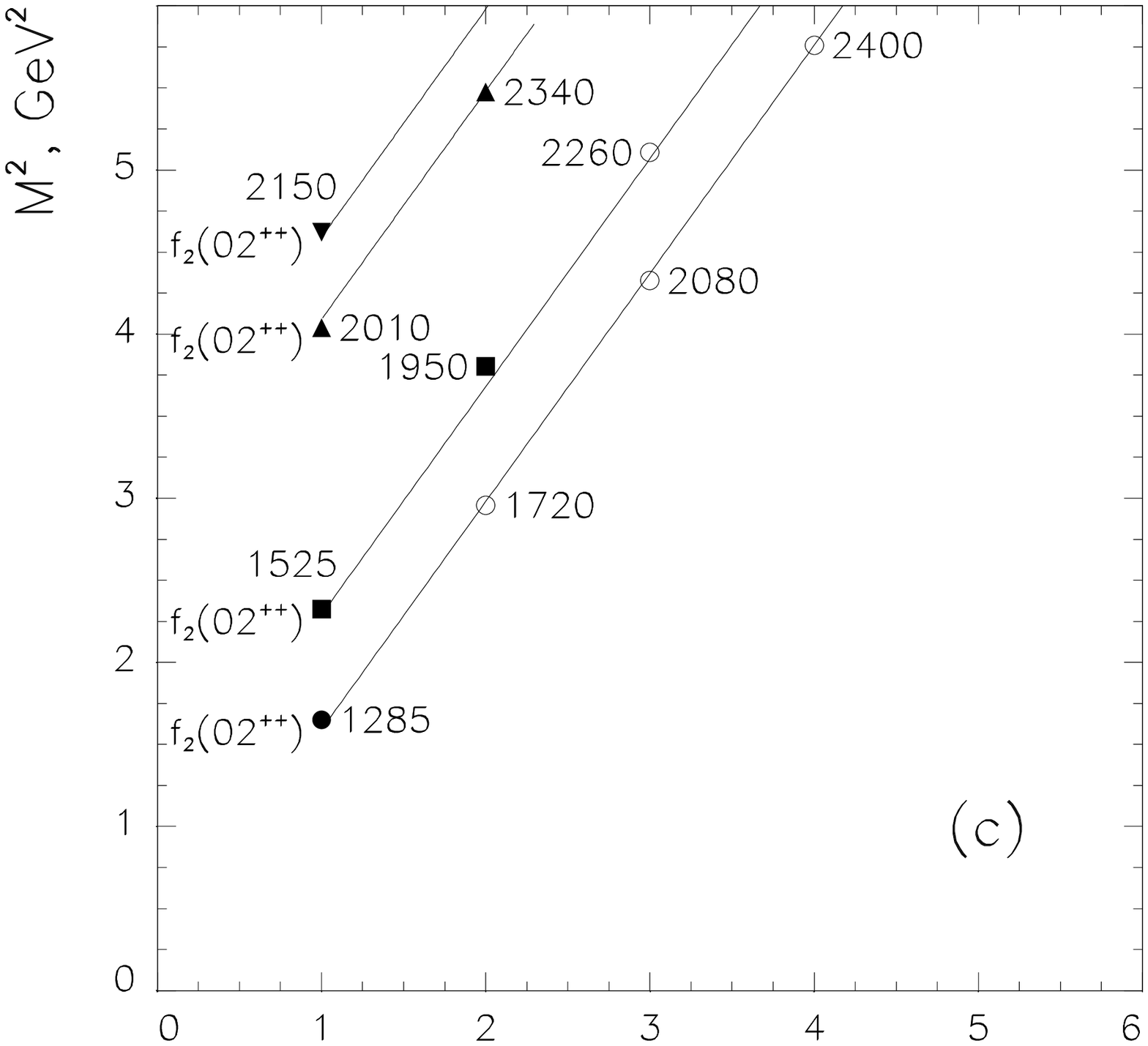,width=8cm}\hspace{-1.5cm}
            \epsfig{file=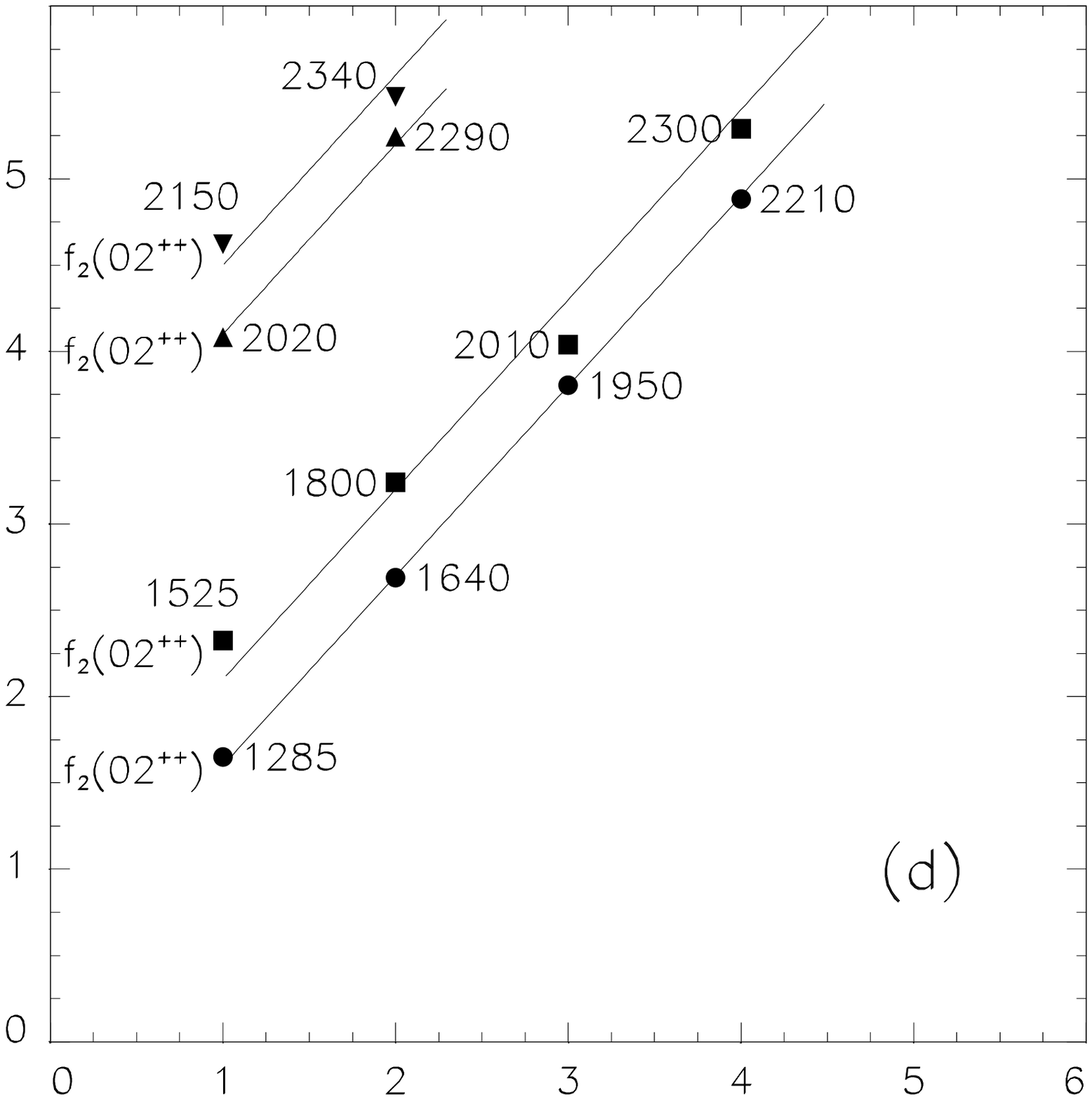,width=8cm}}
\vspace{-1.5cm}
\centerline{\epsfig{file=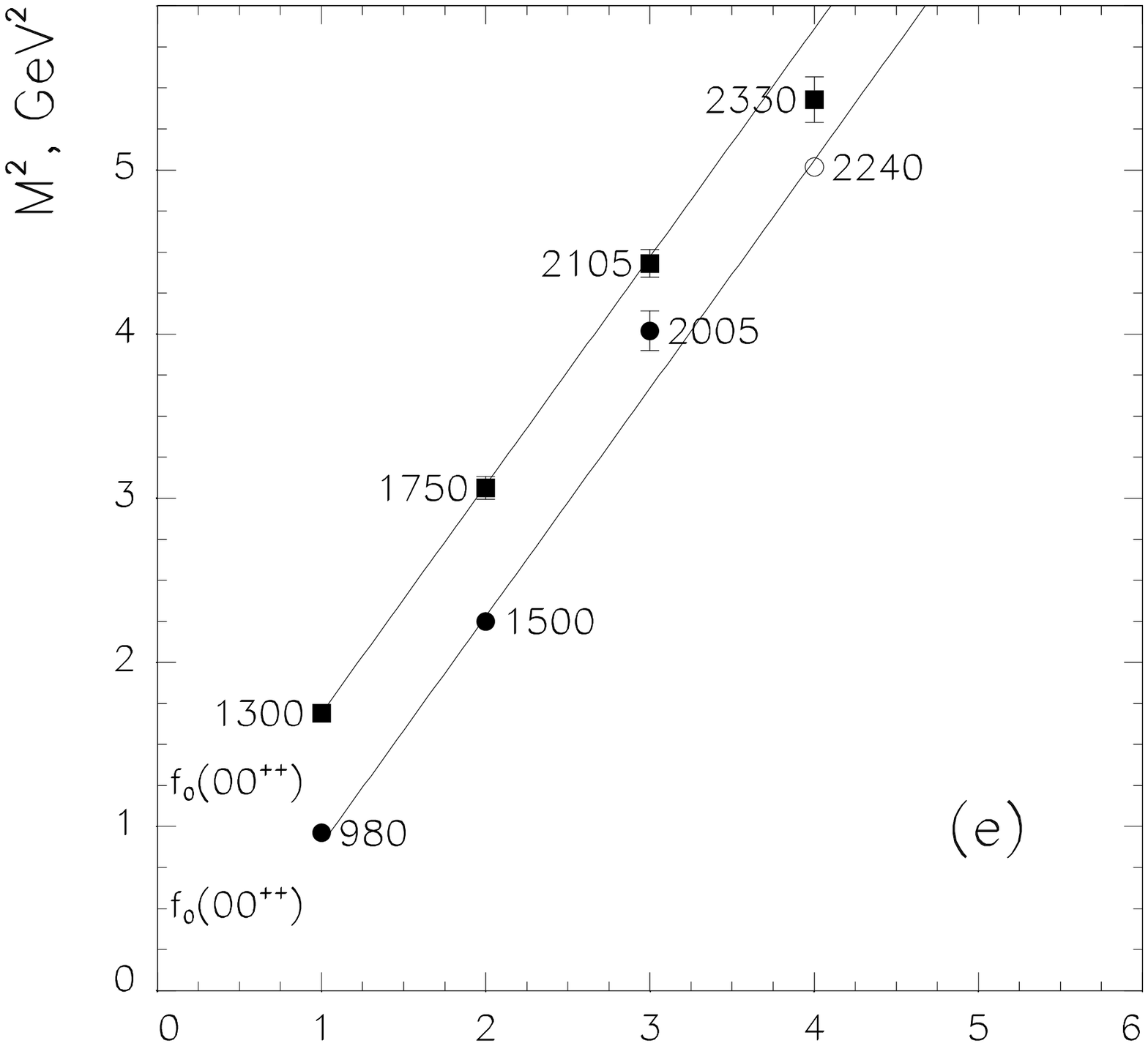,width=8cm}\hspace{-1.5cm}
            \epsfig{file=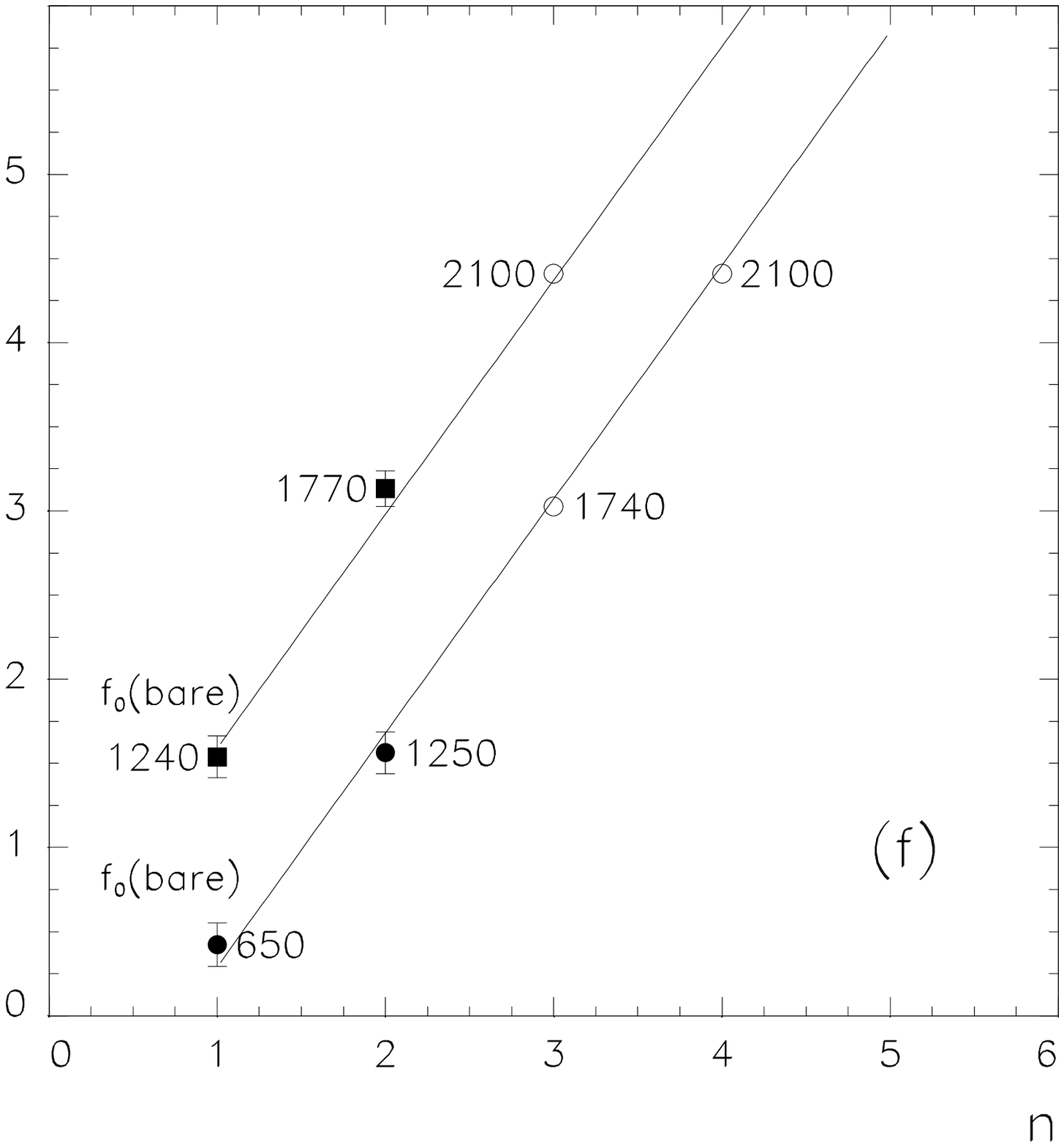,width=8cm}}
\caption{
The $(n, M^2)$-trajectories for the states
 $a_0(10^{++})$, $a_2(12^{++})$ and $a_4(14^{++})$ with a)
$\mu^2 =1.38$ GeV$^2$ and b)  $\mu^2 =1.10$ GeV$^2$;
Two variants for the
$f_2(02^{++})$-trajectories with the slope parameter
 c) $\mu^2= 1.38$ GeV$^2$ and  d) $\mu^2= 1.10$ GeV$^2$.
The $f_0(00^{++})$-trajectories for e) real resonance states
and f) K-matrix pole states.
As in Fig. 1, the
open circles stand for states predicted by the present
classification.}
\end{figure}

\begin{figure}
\centerline{\epsfig{file=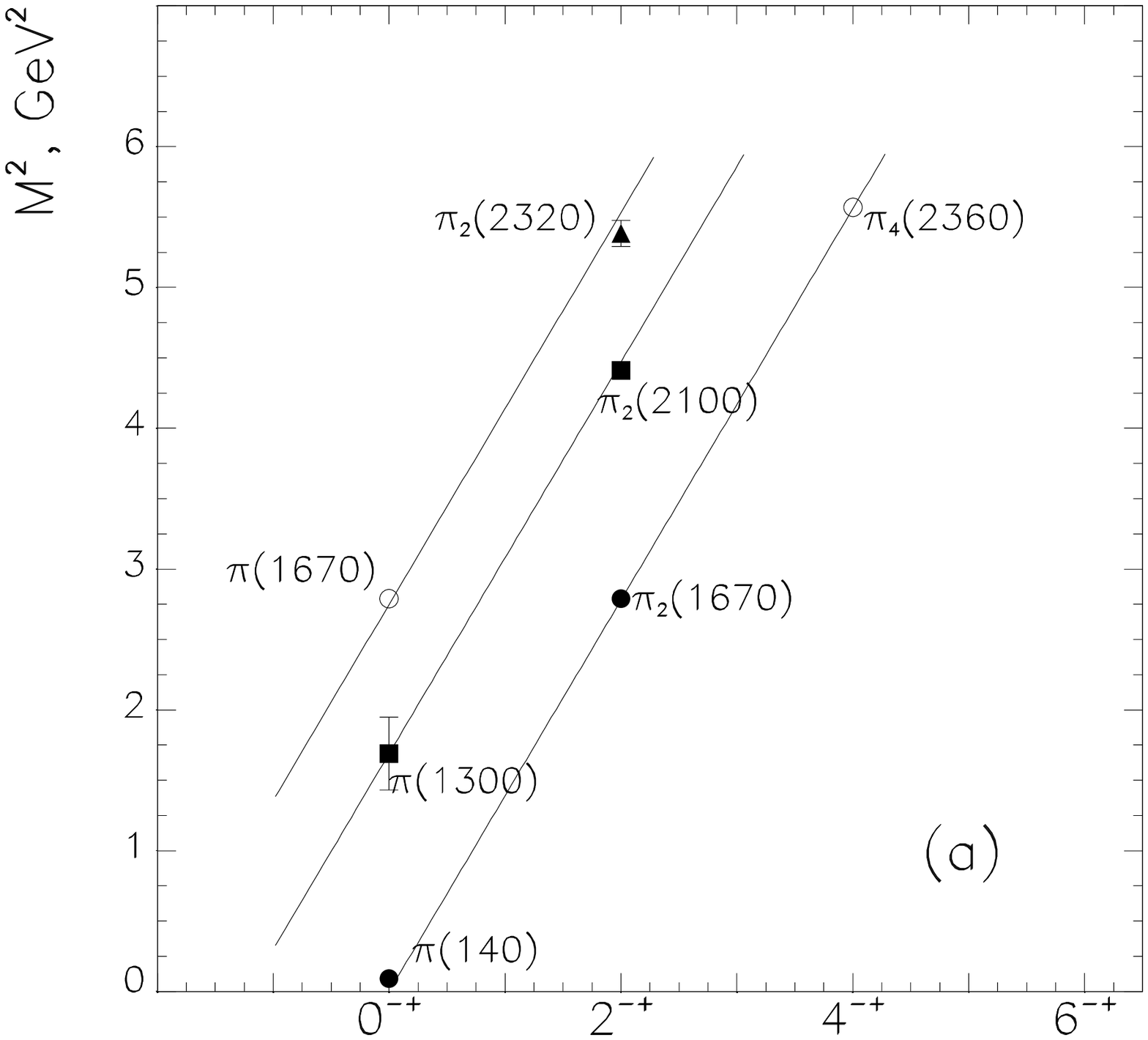,width=8cm}\hspace{-1.5cm}
            \epsfig{file=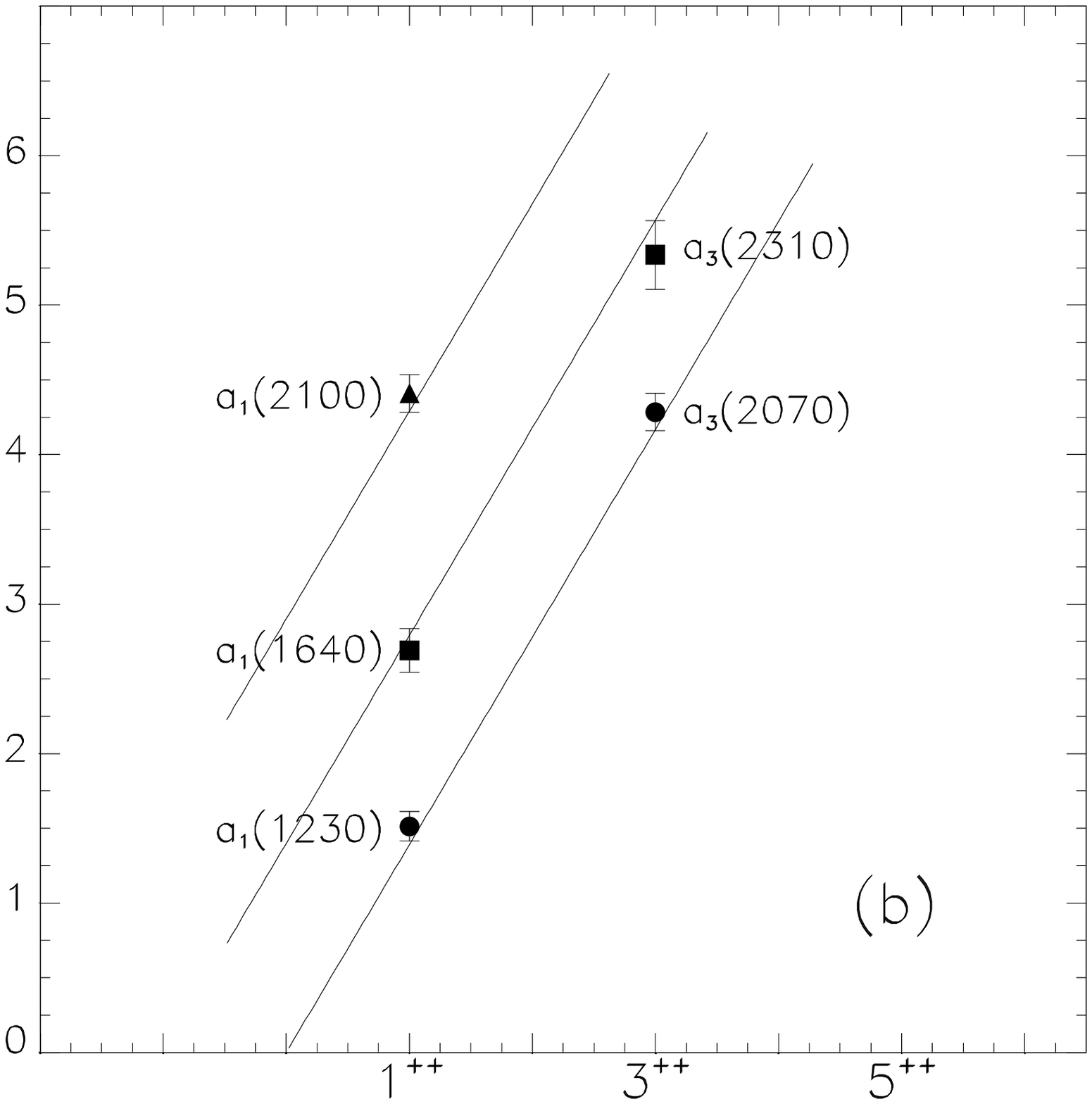,width=8cm}}
\vspace{-1.5cm}
\centerline{\epsfig{file=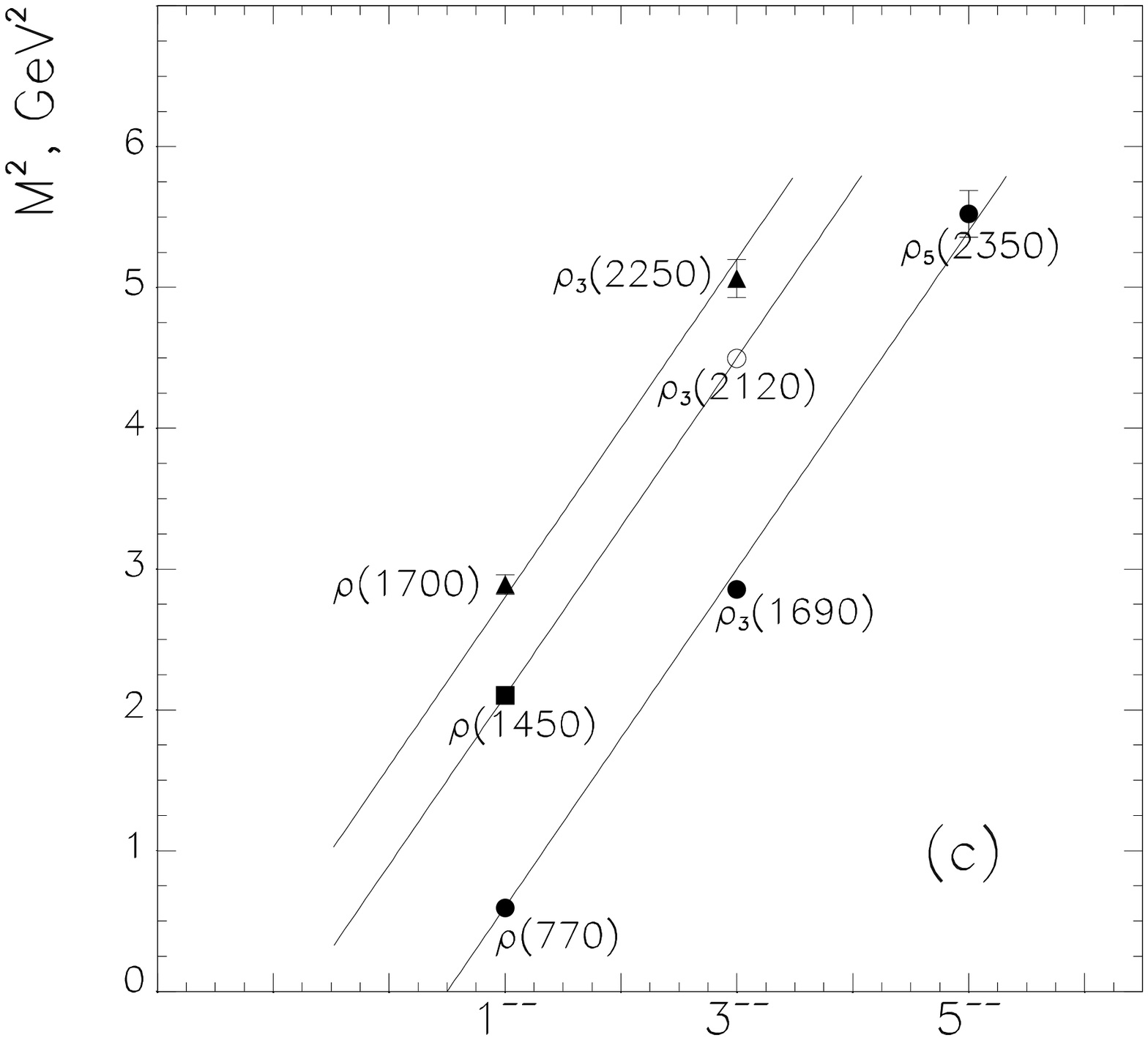,width=8cm}\hspace{-1.5cm}
            \epsfig{file=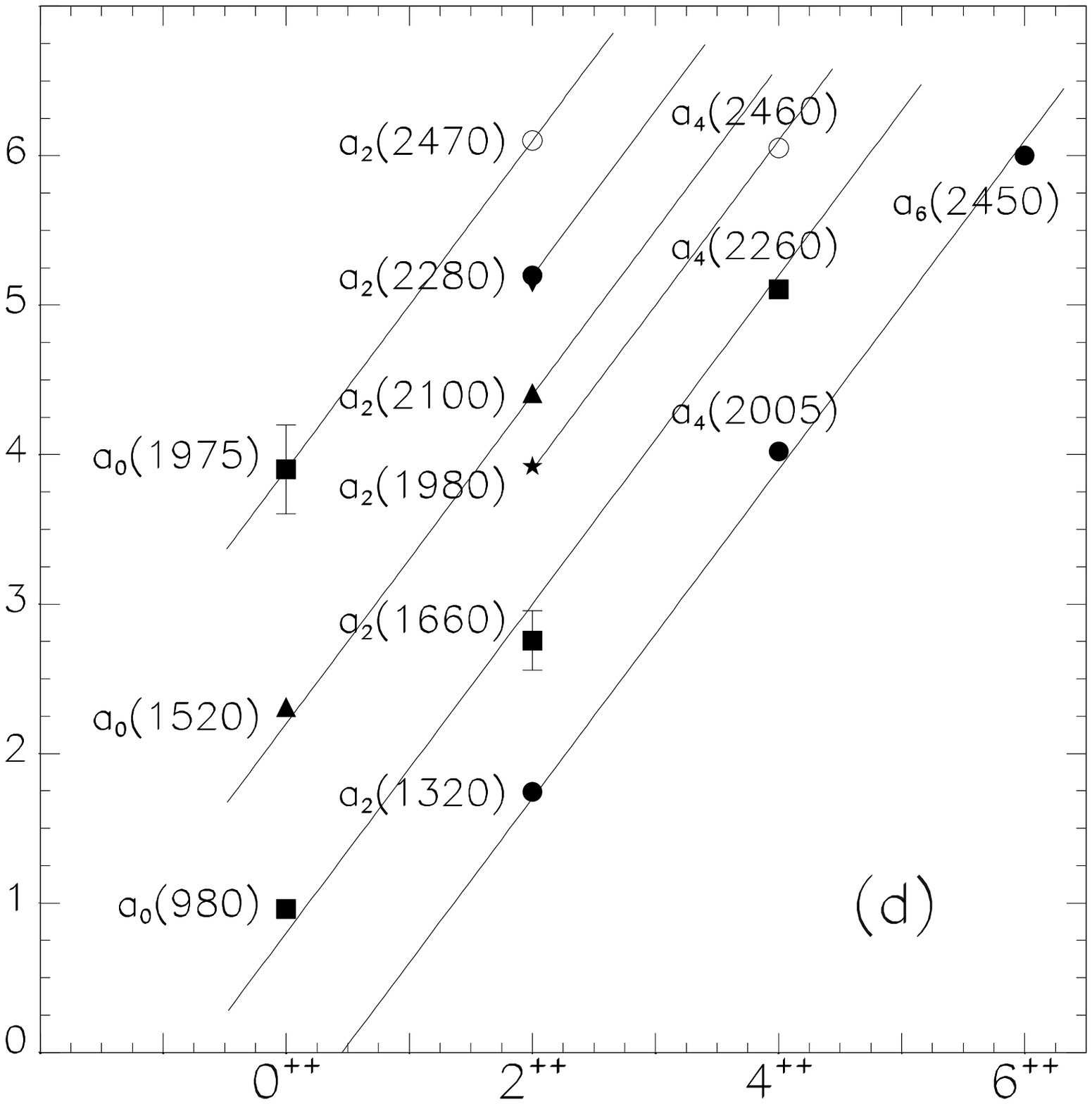,width=8cm}}
\vspace{-1.5cm}
\centerline{\epsfig{file=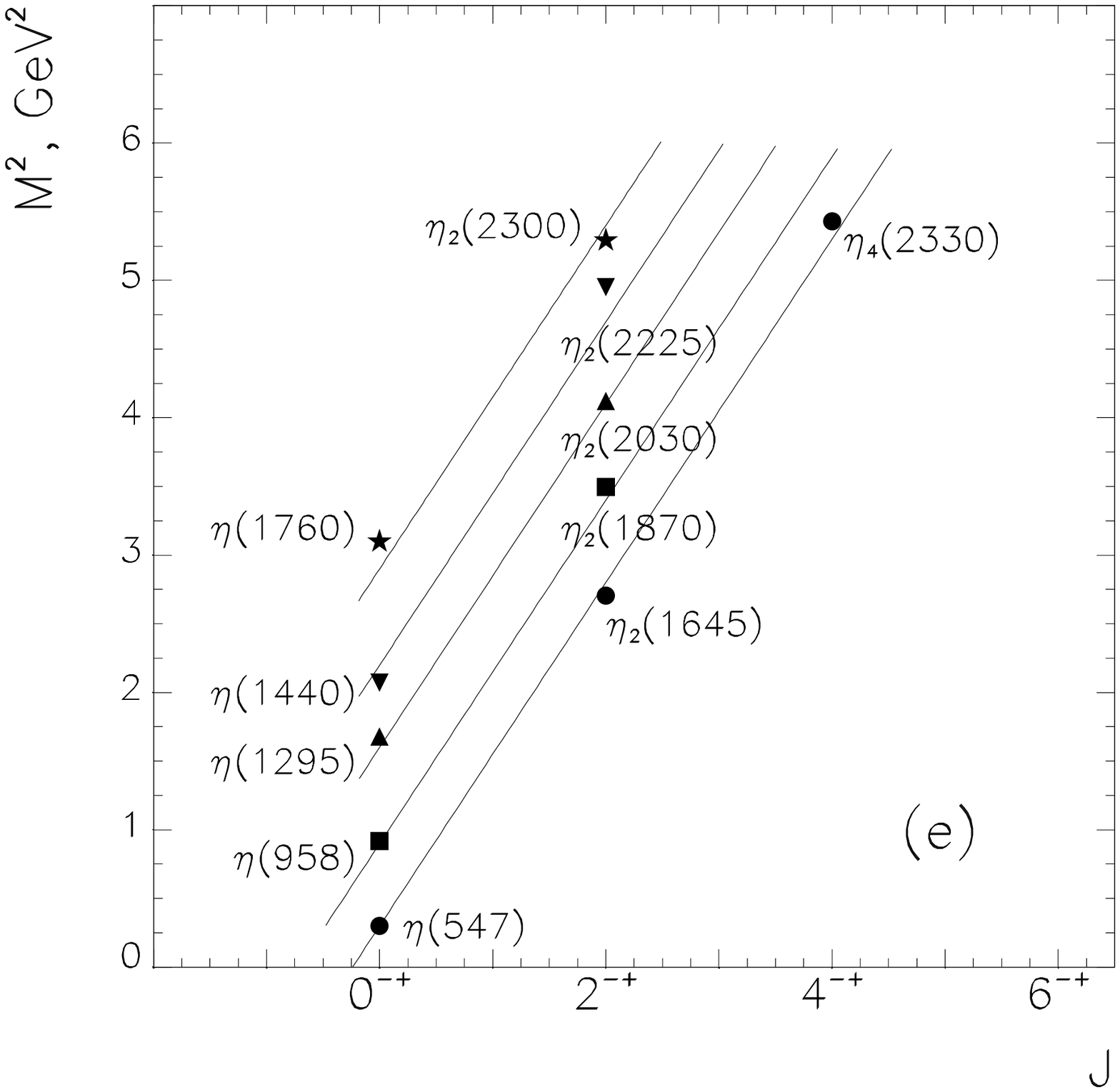,width=8cm}\hspace{-1.5cm}
            \epsfig{file=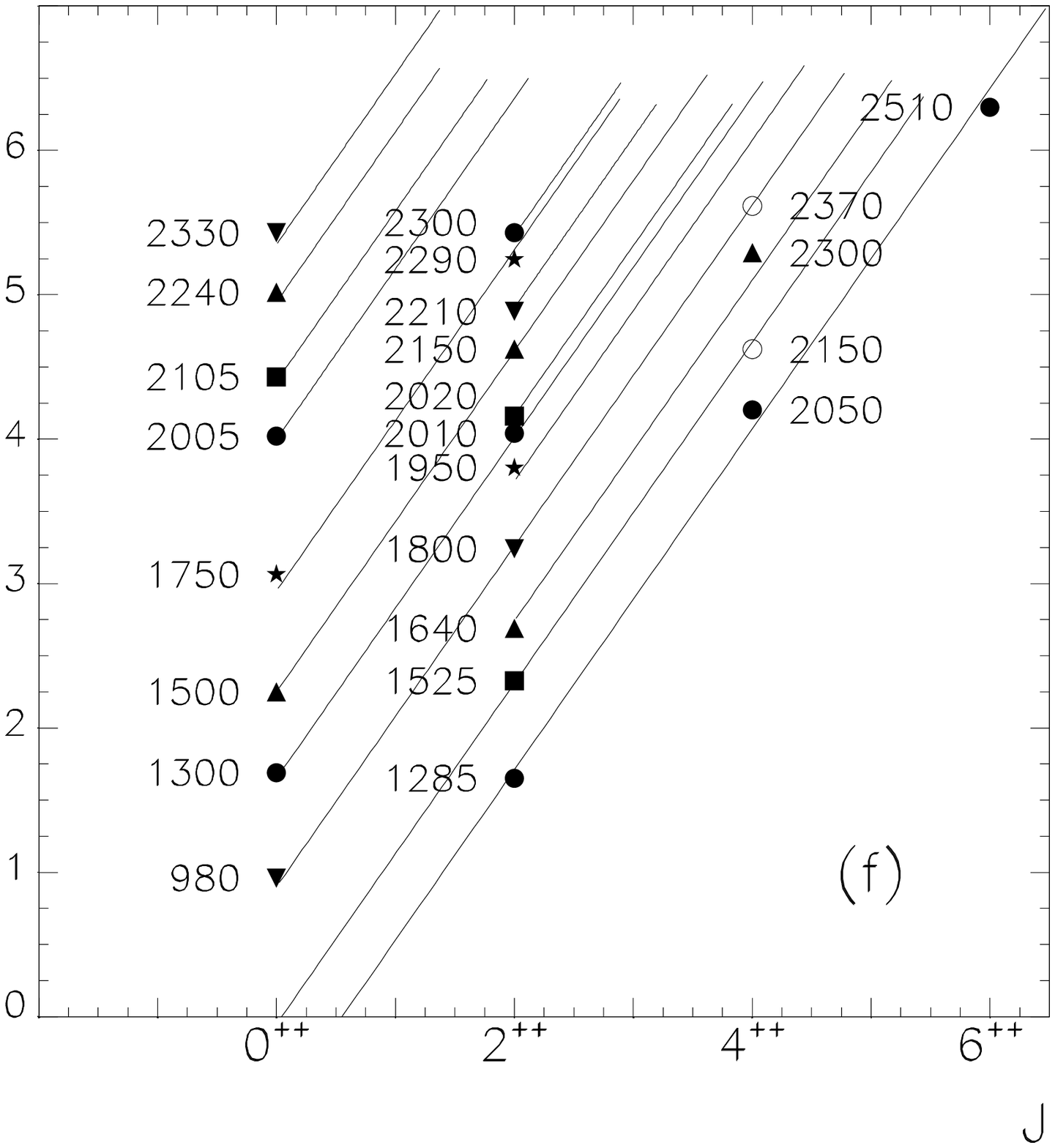,width=8cm}}
\caption{
The $(J, M^2)$-plots for  leading and daugther trajectories:
a) $\pi$-trajectories,
b) $a_1$-trajectories,
c) $\rho$-trajectories,
d) $a_2$-trajectories,
e) $\eta$-trajectories,
f) $P'$-trajectories.}
\end{figure}

\end{document}